\documentclass[10pt,a4paper]{amsart}

\usepackage{amsfonts}
\usepackage{graphicx}
\usepackage{epstopdf}
\usepackage{hyperref}

\ifpdf
  \DeclareGraphicsExtensions{.eps,.pdf,.png,.jpg}
\else
  \DeclareGraphicsExtensions{.eps}
\fi

\usepackage{amssymb}
\usepackage{enumerate}
\usepackage{algorithm}
\usepackage[noend]{algorithmic}

\newcommand{\field}[1][]{\mathbb{F}_{#1}}

\newcommand{\Nset}{\mathbb{N}}

\newcommand{\norm}[3][]{\operatorname{N}^{#1}_{#2}(#3)}

\newcommand{\matrixspace}[2]{{#2}^{#1}}
\newcommand{\tovector}{\mathfrak{v}}

\newcommand{\transpose}[1]{#1^{\mathtt{T}}}

\newcommand{\lclm}[1]{\left[#1\right]_\ell}
\newcommand{\lcrm}[1]{\left[#1\right]_r}
\newcommand{\gcrd}[1]{\left(#1\right)_r}
\newcommand{\gcld}[1]{\left(#1\right)_\ell}
\DeclareMathOperator{\weight}{w}

\theoremstyle{plain}
\newtheorem{theorem}{Theorem}%[section]
\newtheorem*{theorem*}{Theorem}
\newtheorem{proposition}[theorem]{Proposition}

\newtheorem*{corollary*}{Corollary}
\newtheorem{lemma}[theorem]{Lemma}

\theoremstyle{remark}
\newtheorem{remark}[theorem]{Remark}
\newtheorem{example}[theorem]{Example}

\theoremstyle{definition}
\newtheorem{definition}[theorem]{Definition}
\newtheorem{hypothesis}[theorem]{Hypothesis}

\title[Skew differential Goppa codes]{Skew differential Goppa codes and their application to McEliece cryptosystem.}
\thanks{Research funded by grant (PID2019-110525GB-I00 / AEI / 10.13039/501100011033) and partially by the IMAG–Mar\'{\i}a de Maeztu grant
(CEX2020-001105-M / AEI / 10.13039/501100011033)}

\author{Jos\'{e} G\'{o}mez-Torrecillas}
\address{IMAG and Dept. of Algebra, University of Granada.}
\email{gomezj@ugr.es}

\author{F. J. Lobillo}
\address{CITIC and Dept. of Algebra, University of Granada.}\email{jlobillo@ugr.es}

\author{Gabriel Navarro}
\address{CITIC and DECSAI, University of Granada.}
\email{gnavarro@ugr.es}



\begin{document}

\maketitle

\begin{abstract}
A class of linear codes that extends classic Goppa codes to a non-commutative context is defined. An efficient decoding algorithm, based on the solution of a non-commutative key equation, is designed. We show how the parameters of these codes, when the alphabet is a finite field, may be adjusted to propose a McEliece-type cryptosystem. 
  \end{abstract}

\section{Introduction}

Code-based cryptography  proposals  still alive after the Round 4 for the NIST Post-Quantum Cryptography competition.  The strength of these technologies rests upon the hardness of the decoding problem for a general linear code. Of course, an efficient decoding algorithm is required in practice. So, what is already needed is a family of codes with some conveniently masked  properties that allow their efficient decoding. The original McEliece criptosystem took advantage of such  features enjoined by classic Goppa binary codes. 

One way to introduce Goppa codes is the following. Let \(F \subseteq L\) be a field extension and let \(g \in L[x]\) be a polynomial which, in this introduction, we assume irreducible for sake of simplicity. A subset of group of units in \(L[x]/\langle g \rangle\) represented by linear polynomials is selected, and their inverses allow to build a parity check matrix of the Goppa code. The arithmetic in \(L[x]\) is a main tool in the design of efficient decoding algorithms for Goppa codes. 

From an algebraic point of view, our proposal replaces, in the simplest case, the cyclic group of units of \(L[x]/\langle g \rangle\) by a general linear group, whose mathematical structure is more complex. In order to design an efficient decoding algorithm, this non-commutative group is presented as the group of units of Ore polynomials in \(L[x;\sigma,\partial]\) modulo a suitable invariant polynomial \(g\). The arithmetic of this non-commutative polynomial ring is used to design efficient decoding algorithms. Classic Goppa codes are instances of our construction. Therefore, the security of our cryptosystem is expected to be as strong as the original one. 

Section \ref{skewpols} recalls some basic essentials on Ore polynomials and states de definition of skew differential Goppa codes. A non-commutative key equation is derived for these codes (Theorem \ref{thkeyeq}) which turns out to be a left multiple of an equation computed with the help of the Left Euclidean Extended Algorithm in \(L[x;\sigma,\partial]\). 

The topic of Section \ref{decoding} is the design of an efficient decoding algorithm for skew differential Goppa codes. To this end, the positional points are assumed to be P-independent in the sense of \cite{Lam/Leroy:2004}. Under this hypothesis, the non-commutative locator polynomial already locates the error positions  and a decoding algorithm, based upon the solution of the key equation, is provided (Algorithm \ref{Sug}). This algorithm gives a solution in most cases, but it is possible that its output falls in a decoding failure. A subsidiary algorithm solves any of these failures (Algorithm \ref{decodingfailure}). In resume, the combination of both algorithms correctly computes the error added to a codeword up to the correction capability. 

Section \ref{skewGoppaconstruction} describes how to construct parity check matrices and positional points to define skew-Goppa codes suitable to be used in a code-based cryptosystem. The construction is made  in the particular case when the \(\partial = 0\), which makes it transparent. 

The cryptosystem, based on a McEliece scheme with skew Goppa codes over finite fields, is presented in Section \ref{classicalMcEliece}. A discussion on the choice of suitable parameters for the code is included. 

The paper includes two appendices. The first one explains why the linearized Goppa codes introduced in \cite{Wang:2018} are not suitable for our purposes. The second appendix discusses when decoding failure appears. 

There is a patent pending by University of Granada in order to protect some of the results in this work, see \cite{GLN2022patente}.

\section{Skew differential Goppa codes and their non-commutative key equation}\label{skewpols}

In this section, the required algebraic framework is introduced. Let \(\sigma\) be an automorphism of finite order \(\mu\) of a field \(L\).  An additive map \(\partial : L \to L\) is called a \(\sigma\)-derivation if it satisfies \(\partial(ab) = \sigma(a)\partial(b) + \partial(a) b\) for all \(a,b \in L\).  By \(R = L[x;\sigma,\partial]\) we denote the ring of Ore polynomials built from \((\sigma, \partial)\). This is a fundamental example of non-commutative ring,  introduced in \cite{Ore:1933}, whose basic properties may be found in several texts. We follow, and adopt their notation, \cite[Chapter 1, \S 2 and \S 3]{Bueso/Gomez/Verschoren:2003}. In particular, \(R\) is a left and right Euclidean domain. To precise the notation, the left division algorithm computes, given \(f,d\in R\) with \(d\neq 0\), two Ore polynomials \(q, r \in R\) such that \(f = qd + r\) with \(\deg r < \deg d\), where \(\deg\) denotes the degree (in \(x\)) function. We will use the following notation. 
\[
\operatorname{l-quo-rem}(f,d) = (q,r), \qquad \operatorname{l-quo}(f,d) = q.
\]

Given \(f,g \in R\), the notation \(f \mid_r g\) declares that \(f\) is a right divisor of \(g\), which means \(Rg \subseteq Rf\), that is, \(g = uf\) for some \(u \in R\). The notation \(f \mid_\ell g\) is used analogously, meaning \(fR \subseteq gR\). When \((x-\alpha) \mid_r f\), for \(f \in R\) and \(\alpha \in L\), we say that \(\alpha\) is a \emph{right root} of the Ore polynomial \(f\).  Greatest common left/right divisors and least common left/right multiples are well defined since all left/right ideals are principal. Concretely
\[
Rg + Rf = R \gcrd{g,f}, \quad gR + fR = \gcld{g,f}R
\]
and
\[
Rg \cap Rf = R \lclm{g,f}, \quad gR \cap fR = \lcrm{g,f} R. 
\]
There exist Left and Right Extended Euclidean algorithms (LEEA and REEA, for short) that compute 
greatest common divisors and least common multiples on both sides.

For our forthcoming reasoning, we will need a very precise statement of the LEEA which provides the Bezout coefficients in each step of the algorithm. This is given in Algorithm \ref{LEEA_alg}.  

\begin{algorithm}
\caption{\texttt{Left Extended Euclidean  Algorithm}}\label{LEEA_alg}
\begin{algorithmic}
\REQUIRE $f, g \in L[x;\sigma,\partial]$ with $f \neq 0, g \neq 0$.
\ENSURE $\{u_i,v_i,r_i\}_{i=0,\ldots ,h, h+1}$ such that $r_i=u_i f + v_i g$ for every $i$, $r_h=\gcrd{f,g}$,  and $u_{h+1}f = \lclm{f,g}$.
\STATE $r_0\gets f$, $r_1\gets g$.
\STATE $u_0\gets 1$, $u_1\gets 0$.
\STATE $v_0\gets 0$, $v_1\gets 1$.
\STATE $i\gets 1$.
\WHILE{$r_i\not = 0$}
	\STATE find $q_i, r$ such that \(r_{i-1} = q_i r_i + r\) and $\deg r < \deg r_i$ \COMMENT{Left division algorithm}
	\STATE $r_{i+1}\gets r$
	\STATE $u_{i+1}\gets u_{i-1}-q_iu_i$
	\STATE $v_{i+1}\gets v_{i-1}-q_iv_i$
	\STATE $i\gets i+1$
\ENDWHILE
\RETURN $\{u_i,v_i,r_i\}_{i=0,\ldots ,h, h+1}$
\end{algorithmic}
\end{algorithm}

The output of Algorithm \ref{LEEA_alg} enjoys some properties that will be used later. We record them in the following lemma,  whose commutative version may be found in \cite[Lemma 3.8]{VonzurGathen}.

\begin{lemma}\label{LEEA}
Let $f,g\in R$ and $\{u_i,v_i,r_i\}_{i=0,\ldots ,h}$ be the coefficients obtained when applying the LEEA to $f$ and $g$. Then, for all \(i =0, \dots, h\), we have:
\begin{enumerate}
\item $u_i f + v_i g = r_i$.
\item $\gcld{u_i,v_i}=1$.
\item $\deg f=\deg r_{i-1}+\deg v_i$.
\end{enumerate}
\end{lemma}
\begin{proof}
The proof given in \cite[Lemma 24]{gln2017sugiyama} works here step by step. 
\end{proof}

\begin{remark}\label{lclmandgcld}
Let \(f,g, f', g'  \in R\) nonzero Ore polynomials such that \(f'f = g'g\). Then \(\lclm{f,g} = f'f\) if and only if \(\gcld{f',g'} = 1\). In fact, assume \(\lclm{f,g} = f'f = g'g\) and let \(d = \gcld{f',g'}\). Then \(f' = d f''\) and \(g' = d g''\). Hence \(d f'' f = d g'' g\), so \(\lclm{f,g} \mid_r f'' f = g'' g\). Therefore \(f'f \mid_r f''f\), i.e., \(f' \mid_r f''\). It follows that  \(d\) is a unit, so \(d \in L\) and \(\gcld{f',g'} = 1\). Conversely, assume \(\gcld{f',g'} = 1\). If \(\lclm{f,g} = f''' f = g''' g\), since \(\lclm{f,g} \mid_r f'f = g'g\) there exists \(c\in R\) such that \(c \lclm{f,g} = f'f = g'g\). It follows \(c f''' = f'\) and \(c g''' = g'\), hence \(c \mid_l \gcld{f',g'}\). Therefore \(c \in L\) and \(\lclm{f,g} = f'f = g'g\).

The analogous result for least common right multiples and greatest common right divisors also holds. 
\end{remark}

Let \(0 \neq g \in R\) be invariant, i.e. \(Rg = gR\). Therefore, \(R/Rg\) is a ring. It is easy to check that \(fh = g\) if and only if \(h'f = g\), where \(gh = h'g\), so \(f \mid_r g\) if and only if \(f \mid_l g\). In particular, since \(x-\gamma\) is irreducible for all \(\gamma \in L\), \(\gcrd{x-\gamma,g} = 1\) if and only if \(\gcld{x-\gamma,g} = 1\).

Observe that \(\gcld{x-\gamma,g} = 1\) means that \(x-\gamma + Rg\) is a unit in \(R/Rg\), so there exists a unique \(h \in R\) with \(\deg(h) < \deg(g)\) such that
\((x-\gamma)h - 1 \in Rg\) and \(h(x-\gamma) -1 \in Rg\).

\begin{definition}\label{sdGdef}
Let \(F \subseteq L\) be a field extension. Let \(g \in R = L[x;\sigma,\partial]\) be a nonzero invariant polynomial. Let \(\alpha_0, \dots, \alpha_{n-1} \in L\) be different elements such that \(\gcrd{x - \alpha_i, g} = 1\) for all \(0 \leq i \leq n-1\), let \(h_i \in R\) such that \(\deg(h_i) < \deg(g)\) and 
\begin{equation}\label{eq:rootsinverses}
(x - \alpha_i) h_i - 1 \in R g,
\end{equation}
and let \(\eta_0, \dots, \eta_{n-1} \in L^*\). A (generalized) skew differential Goppa code \(\mathcal{C} \subseteq F^n\) is the set of vectors \((c_0,\dots, c_{n-1}) \in F^n\) such that 
\begin{equation}\label{eq:Goppadef}
\sum_{i=0}^{n-1} h_i \eta_i c_i = 0.
\end{equation}
By a degree argument, \eqref{eq:Goppadef} is equivalent to
\begin{equation}\label{eq:Goppadef2}
\sum_{i=0}^{n-1} h_i \eta_i c_i \in R g.
\end{equation}
We say that \(\{\alpha_0, \dots, \alpha_{n-1}\}\) are the positional points, \(g\) is the (skew differential) Goppa polynomial and \(h_0, \dots, h_{n-1}\) are the parity check polynomials. If \(\partial = 0\), we just call it a (generalized) skew Goppa code. 
\end{definition}

\begin{remark}
A classic Goppa code is an instance of skew differential Goppa codes when \(\sigma\) is the identity map, \(\partial = 0\) and \(\eta_i = 1\) for all \(0 \leq i \leq n-1\). 
\end{remark}

\begin{remark}
In \cite{Wang:2018}, linearized Goppa codes are introduced. Since the ring of linearized polynomials over finite field is isomorphic to the ring of Ore polynomials built from the Frobenius automorphism with trivial skew derivation, linearized Goppa codes become instances of skew differential Goppa codes.  However, the material from \cite{Wang:2018} is not suitable for our purposes, as discussed in Appendix \ref{Wang}. 
\end{remark}

For the rest of this section a skew differential Goppa code \(\mathcal{C}\) is fixed. Let \(\{\varepsilon_i ~|~ 0 \leq i \leq n-1\}\) be the canonical basis of \(F^n\). Assume \(c \in \mathcal{C}\) is transmitted and \(r \in F^n\) is received. Therefore
\[
r = c + e
\]
for some \(e = \sum_{j=1}^\nu e_{j} \varepsilon_{k_j}\) with \(e_j \neq 0\) for \(1 \leq j \leq \nu\). The \emph{syndrome polynomial} is defined and computed as 
\[
s = \sum_{i=0}^{n-1} h_i \eta_i r_i. 
\]
By \eqref{eq:Goppadef}, it follows that 
\begin{equation}\label{eq:syndromeanderrors}
s - \sum_{j=1}^\nu h_{k_j} \eta_{k_j} e_j = \sum_{i=0}^{n-1} h_i \eta_i c_i \in Rg = gR.
\end{equation}

We define the (non-commutative) \emph{error locator polynomial} as 
\[
\lambda = \lclm{\{x - \alpha_{k_j}~|~1 \leq j \leq \nu\}} \in R.
\]
Then \(\deg(\lambda) \leq \nu\) and, for each \(1 \leq j \leq \nu\), there exists \(\rho_{k_j} \in R\) such that \(\deg(\rho_{k_j}) < \nu\) and 
\begin{equation}\label{eq:rho's}
\lambda = \rho_{k_j} (x-\alpha_{k_j}).
\end{equation}
The \emph{error evaluator polynomial} is defined as 
\[
\omega = \sum_{j=1}^\nu \rho_{k_j} \eta_{k_j} e_j.
\]
It follows that \(\deg(\omega) < \nu\). 

Our next aim is to derive and solve a non-commutative key equation that relates syndrome, locator and evaluator polynomials. The solution will require the following lemma.

\begin{lemma}\label{keyeqsolver}
Let \(f,g \in R\) such that \(\deg f < \deg g = \chi\). Assume that there exist \(\kappa, \lambda, \omega \in R\) such that 
\(
\kappa g + \lambda f = \omega, \deg \lambda \leq \left\lfloor \frac{\chi}{2} \right\rfloor, \deg \omega <\left\lfloor \frac{\chi}{2} \right\rfloor.
\)
Let $u_I,v_I$ and $r_I$ be the (partial) Bezout coefficients returned by the LEEA with input $g$ and $f$, where $I$ is the index determined by the conditions $\deg r_{I-1} \geq \left\lfloor \frac{\chi}{2} \right\rfloor$ and $\deg r_I < \left\lfloor \frac{\chi}{2} \right\rfloor$. Then there exists \(h \in R\) such that \(\kappa = h u_I\), $\lambda = h v_I$ and $\omega = h r_I$.
\end{lemma}

\begin{proof}
Since \(\kappa g + \lambda f = \omega\), \(\deg \lambda \leq \left\lfloor \frac{\chi}{2} \right\rfloor\) and \(\deg \omega < \left\lfloor \frac{\chi}{2} \right\rfloor\), it follows that $\deg \kappa < \left\lfloor \frac{\chi}{2} \right\rfloor$. By Lemma \ref{LEEA}, $\deg v_I + \deg r_{I-1} = \chi$, so that $\deg v_I \leq \chi - \left\lfloor \frac{\chi}{2} \right\rfloor$.

Write $\lclm{\lambda,v_I} = a \lambda = b v_I$, where $a,b\in R$ with $\deg a \leq \deg v_I \leq \chi - \left\lfloor \frac{\chi}{2} \right\rfloor$ and $\deg b \leq \deg \lambda \leq \left\lfloor \frac{\chi}{2} \right\rfloor$. Then $\gcld{a,b}=1$ by Remark \ref{lclmandgcld}. 

From \(\kappa g + \lambda f = \omega\) we get
 \begin{equation}\label{keyeq2} 
a \kappa g + a \lambda s = a \omega. 
\end{equation}
By Lemma \ref{LEEA}, we have \(u_I g + v_I f = r_I\), which we multiply on the left by $b$ to get
\begin{equation}\label{euclideq2}
b u_I g + b v_I s = b r_I.
\end{equation}
Hence, from \eqref{keyeq2} and \eqref{euclideq2}, 
\begin{equation}\label{key-euclid}
(a \kappa - b u_I) g = a \omega - b r_I.
\end{equation} 
Since 
\begin{multline*}
\deg(a \omega - b r_I) \leq \max\left\{ \deg a + \deg \omega, \deg b + \deg r_I \right\} \\
< \max\left\{\chi - \left\lfloor \frac{\chi}{2} \right\rfloor + \left\lfloor \frac{\chi}{2} \right\rfloor, \left\lfloor \frac{\chi}{2} \right\rfloor + \left\lfloor \frac{\chi}{2} \right\rfloor\right\} = \chi = \deg g,
\end{multline*}
it follows, from \eqref{key-euclid}, that $a \kappa = b u_I$ and $a \omega = b r_I$. Actually, $\gcld{a,b}=1$ yields $\lclm{\kappa,u_I} = a \kappa = b u_I$ and \(\lclm{\omega,r_I} = a \omega = b r_I\) by Remark \ref{lclmandgcld}. In particular, $\deg a \leq \deg r_I < \left\lfloor \frac{\chi}{2} \right\rfloor$.

Let $\lcrm{a,b} = a a' = b b'$. Since $\lclm{\lambda,v_I}$ is a right multiple of $a$ and $b$, there exists $m \in R$ such that $\lclm{\lambda,v_I} = \lcrm{a,b} m$. Then $a \lambda = b v_I = a a' m = b b' m$. Thus, $\lambda = a' m$ and $v_I = b' m$ and, by minimality, $\gcrd{\lambda,v_I} = m$. Similar arguments prove that there exist $m',m''\in R$ such that $u_I = b' m'$ and $\kappa = a' m'$, and that $r_I = b' m''$ and $\omega = a' m''$. Nevertheless, by Lemma \ref{LEEA}, $\gcld{u_I,v_I} = 1$, so $b'=1$. In this way, $b = a a'$ and we get $\lambda = a' v_I$, $\omega = a' r_I$ and $\kappa = a' u_I$. This completes the proof.
\end{proof}

\begin{theorem}\label{thkeyeq}
The error locator \(\lambda\) and the error evaluator \(\omega\) polynomials satisfy the non-commutative key equation
\begin{equation}\label{keyeqmod}
\omega = \kappa g + \lambda s, 
\end{equation}
for some \(\kappa \in R\). 
Assume that \(\nu \leq t = \left \lfloor \frac{\deg g}{2} \right\rfloor\). Let $u_I,v_I$ and $r_I$ be the Bezout coefficients returned by the left extended Euclidean algorithm with input $g$ and $s$, where $I$ is the index determined by the conditions $\deg r_{I-1}\geq t$ and $\deg r_I<t$. Then there exists \(h \in R\) such that \(\kappa = h u_I\), $\lambda = h v_I$ and $\omega = h r_I$.
\end{theorem}

\begin{proof}
Since $(x - \alpha_i) h_i  + R g  = 1 + Rg$ for all \(0 \leq i \leq n-1\), we get from \eqref{eq:syndromeanderrors} the following computation in the ring $R/Rg$: 
\begin{equation*}
\begin{split}
\lambda s + Rg &=  \sum_{j=1}^\nu \lambda h_{k_j} \eta_{k_j} e_j + R g \\
&= \sum_{j=1}^\nu \rho_{k_j} (x-\alpha_{k_j}) h_{k_j} \eta_{k_j} e_j + R g \\
&= \sum_{j=1}^\nu \rho_{k_j} \eta_{k_j} e_j + R g \\
&= \omega + R g.
\end{split}
\end{equation*}
This proves \eqref{keyeqmod}. By construction, 
\[
\deg s \leq \max\left\{ \deg(h_i) | 0 \leq i \leq n-1 \right\} < \deg g, 
\]
so the second statement of the theorem follows from Lemma \ref{keyeqsolver}.
\end{proof}

\section{Decoding algorithms}\label{decoding}

From now on we assume \[\nu \leq t = \left\lfloor \frac{\deg g}{2} \right\rfloor.\] It follows from Theorem \ref{thkeyeq} that the condition \(\gcld{\lambda,\omega} = 1\) implies that \(\lambda\) and \(\omega\) are left associated to \(v_I\) and \(r_I\), respectively. Hence, under this condition the LEEA computes the locator and evaluator polynomials. In the commutative case, it is easy to check that locator and evaluator are always relatively prime. Although, in our experiments, most of examples  in our current non-commutative setting  already satisfy the condition \(\gcld{\lambda,\omega} = 1\), this is not always the case (see Example \ref{criterion}). For a correct decoding, we need to know when \(v_I,r_I\) are actually the locator and the evaluator polynomials and if the locator already locates the error positions.

In order to proceed, we need to use the notion of a left \(P\)--independent set in the sense of  \cite{Lam/Leroy:1988b, Delenclos/Leroy:2007}. From now on, we assume the following hypothesis on the positional points. 

\begin{hypothesis}\label{independence}
We assume that \(\{\alpha_0, \dots, \alpha_{n-1}\} \subseteq L\) is left P-independent, that is, 
 \begin{equation}\label{eq:leftPindependent}
\deg \lclm{\{x - \alpha_i ~|~0 \leq i \leq n-1\}} = n.
\end{equation}\end{hypothesis}

As a consequence of Hypothesis \ref{independence}, \(\deg(\lambda) = \nu\). Let us deduce that \(\lambda\) already locates the error positions.

\begin{proposition}\label{rootslocator}
\(x- \alpha_k \mid_r \lambda\) if and only if \(k \in \{k_1, \dots, k_\nu\}\).
\end{proposition}

\begin{proof}
If \(x-\alpha_k \mid_r \lambda\) with  \(k \notin \{k_1, \dots, k_\nu\}\) then the set \(\{\alpha_k, \alpha_{k_1}, \dots, \alpha_{k_\nu}\}\) is left P-dependent. 
\end{proof}

\subsection{Decoding algorithm with unlikely decoding failure}
We will give in Proposition \ref{criterion} a criterion on the partial outputs of LEEA to decide  if  \(\lambda \) is left associated to \(v_I\). This leads to a decoding algorithm  (Algorithm \ref{Sug}) that turns out to work most cases. Later, we will discuss how to correctly decoding in  the rare cases when decoding failure happens in the output of Algorithm \ref{Sug}. Our approach is adapted from \cite[Lemma 26 and Theorem 15]{gln2017sugiyama}.

\begin{lemma}\label{locatorandp_j}
Let $\{i_1, \dots, i_m\}\subseteq \{0,\ldots ,n-1\}$ with $1 < m \leq n$, and \[ f=\lclm{x - \alpha_{i_1}, \ldots ,x - \alpha_{i_m}}.\] Let $f_1,\ldots, f_m\in R$ such that $f = f_j (x - \alpha_{i_j})$ for all $1 \leq j\leq m$. Then:
\begin{enumerate}
\item $\lcrm{f_1, \ldots, f_m} = f$ and $\gcld{f_1, \ldots, f_m} = 1$. 
\item  \(R/f R = \bigoplus_{j=1}^m f_j R/f R\).
\item For any $h\in R$ with $\deg h < m$ there exist $a_1,\ldots ,a_m \in L$ such that $h = \sum_{j=1}^m f_j a_j$.
\item The set \(\{f_1, \dots, f_m\}\) gives, modulo $f R$, a basis of \(R/f R\) as an $L$-vector space.
\end{enumerate}
\end{lemma}

\begin{proof}
(1) By Hypothesis \ref{independence}, \(\{\alpha_{i_1}, \dots, \alpha_{i_m}\}\) is left P-independent. So, by \eqref{eq:leftPindependent}, $\deg f = m$ and, thus, $\deg f_j = m-1$ for every $j=1,\ldots ,m$.  Since $m > 1$,  the degree of $\lcrm{f_1, \dots, f_m}$ must be at least $m-1+1 = m$. But $f$ is obviously a common left multiple of $f_1, \dots, f_m$, whence $f = \lclm{f_1, \dots, f_m}$. It is straightforward to check that \(\gcld{f_1, \ldots, f_m} = 1\), otherwise there would be a left common multiple of \(x - \alpha_{i_j}\) for \(1 \leq j \leq m\) with degree smaller than \(\deg f\). 

(2) Since \(f R \subseteq f_j R\) for all \(1 \leq j \leq m\) and $\gcld{f_1, \ldots, f_m} = 1$, we get \(R/fR = \sum_{j=1}^m f_j R / f R\). Observe that \(f_j R/f R \cong R/(x - \alpha_{i_j})R\) is one-dimensional over $L$. Since the dimension of $R/fR$ as an $L$--vector space is $\deg f = m$, we get that the sum is direct.

(3) and (4) follow from (2).
\end{proof}

\begin{proposition}\label{criterion}
Let $u, v, r \in R$ such that $u g + v s = r$, $h u = \kappa$, $h v = \lambda$ and $h r = \omega$ for some $h \in R$. Let $T=\{l_1, l_2, \dots,l_m\} = \{ 0 \leq l \leq n-1 ~|~ (x-\alpha_l) \mid_r v\}$. Then $m = \deg v$ if and only if $\deg h = 0$.
\end{proposition}

\begin{proof}
Since \(v \mid_r \lambda\), every right root of \(v\) is a right root of \(\lambda\), hence \(\{l_1, \dots, l_m\} \subseteq \{k_1, \dots, k_\nu\}\) by Proposition \ref{rootslocator}. We reorder the set of error positions in such a way that $T=\{k_1, \ldots ,k_m\}$ with $m \leq \nu$. If $\deg h = 0$, then $m = \nu$ and $\deg v = \nu$ by \eqref{eq:leftPindependent}, since \(\{\alpha_{k_1}, \dots, \alpha_{k_\nu}\}\) is left P-independent. Conversely, if $m = \deg v$, then 
\[
v = \lclm{\{x - \alpha_{k_j} ~|~1 \leq j \leq m\}}
\]
by \eqref{eq:leftPindependent} and Hypothesis \ref{independence}. Recall that \(\lambda = \rho_{k_j} (x - \alpha_{k_j})\) and write $v = \rho'_j (x - \alpha_{k_j})$ for all $1 \leq j \leq m$. Since 
\[
\deg r = \deg \omega - \deg h = \deg \omega + \deg v - \deg \lambda \leq \nu-1+m-\nu = m-1,
\] 
we get from Lemma \ref{locatorandp_j} that $r = \sum_{i=1}^m \rho'_i a_i$ for some $a_{1},\ldots, a_{m} \in L$. On the other hand, $\lambda = h v$. Thus, for any $1 \leq j \leq m$, $\rho_{k_j} (x - \alpha_{k_j}) = h \rho'_j (x - \alpha_{k_j})$, so $\rho_{k_j} = h \rho'_j$. Now, $h r = \omega$, so 
\begin{equation}\label{aux}
\sum_{j=1}^m \rho_{k_j} a_j = h \left( \sum_{j=1}^m \rho'_j a_j \right) = h r = \omega = \sum_{j=1}^m \rho_{k_j} e_j + \sum_{j=m+1}^\nu \rho_{k_j} e_{j}.
\end{equation}
By Lemma \ref{locatorandp_j}, $\{\rho_{k_1}, \ldots, \rho_{k_\nu}\}$ is a basis of $R/\lambda R$ as a right $L$--vector space. Therefore, since $e_j \neq 0$ for every $1 \leq j \leq \nu$, equation \eqref{aux} implies that $m = \nu$ and, thus, $\deg h = 0$.
\end{proof}

Theorem \ref{thkeyeq} and Proposition \ref{criterion} support the consistence of a decoding algorithm which is presented as Algorithm \ref{Sug}.

\begin{algorithm}[h]
\caption{Decoding algorithm for skew differential Goppa codes with unlikely decoding failure}\label{Sug}
\begin{algorithmic}[5]
\REQUIRE A skew differential Goppa code \(\mathcal{C}\) of length \(n\), correction capability \(t\), positional points \(\{\alpha_0, \dots, \alpha_{n-1}\} \subseteq L\), \(\eta_0, \dots, \eta_{n-1} \in L^*\), skew differential Goppa invariant polynomial \(g \in L[x;\sigma,\partial]\) with \(\deg(g) = 2t\), and parity check polynomials \(h_0, \dots, h_{n-1} \in L[x;\sigma,\partial]\) of degree \(2t-1\).
\REQUIRE A received polynomial $y=\sum_{i=0}^{n-1}y_ix^i$.
\ENSURE A vector \(e \in F^n\) such that \(\weight(e) \leq t\) and \(y-e \in \mathcal{C}\), or \emph{decoding failure}.
\STATE \(s \gets \sum_{i=0}^{n-1} h_i \eta_i y_i\)
\IF{$s=0$}
	\RETURN $y$
\ENDIF
\STATE \COMMENT{LEEA}
\STATE \(r_{prev} \gets g\), \(r_{curr} \gets s\), \(v_{prev} \gets 0\), \(v_{curr} \gets 1\)
\WHILE{\(\deg(r_{curr}) \geq t\)}
	\STATE \(f,r \gets \operatorname{l-quo-rem}(r_{prev},r_{curr})\)
	\STATE \(v \gets v_{prev} - f v_{curr}\), \(v_{prev} \gets v_{curr}\), \(r_{prev} \gets r_{curr}\), \(v_{curr} \gets v\), \(r_{curr} \gets r\)
\ENDWHILE
\STATE \COMMENT{Finding error positions}
\STATE $pos \gets \{\}$, \(other = \{0, \dots, n-1\}\)
\FOR{\(0 \leq i \leq n-1\)}
	\IF{$\alpha_i$ is a right root of $v_{curr}$}
		\STATE $pos \gets pos \cup \{i\}$, \(other = other \setminus \{i\}\)
	\ENDIF
\ENDFOR
\IF{$\deg(v_{curr}) > |pos|$}
	\STATE `Decoding failure'
	\STATE \textbf{stop}
\ENDIF
\STATE \COMMENT{Finding error values}
\FOR{\(j \in pos\)}
	\STATE $\rho_j\gets \operatorname{l-quo}(v_{curr},x-\alpha_j)$
\ENDFOR
\STATE Solve the linear system $r_{curr}=\sum_{j\in pos}\rho_j \eta_j e_j$ \label{Sug:LS}
\STATE $e \gets \sum_{j\in pos} e_jx^j$
\RETURN $e$
\end{algorithmic}
\end{algorithm}

\subsection{Solving decoding failures}

Proposition \ref{criterion} gives a sufficient condition which tells us if we have actually found the solution of \eqref{keyeqmod}, and, therefore, the output of Algorithm \ref{Sug} is the error polynomial. Otherwise, we have not found the locator, so we need to find new right roots of \(\lambda\).

\begin{proposition}\label{addingnewroot}
Let $u, v, r \in R$ such that $u g + v s = r$, $h u = \kappa$, $h v = \lambda$ and $h r = \omega$ for some $h \in R$. Let \(k \in \{0, \dots, n-1\}\) such that \(x-\alpha_k \nmid_r v\) but \(x - \alpha_k \mid_r \lambda\).  Set  \(v' = \lclm{x-\alpha_k,v}\) and let \(h'' \in R\) such that \(h'' v = v'\). Define  \(u' = h'' u\) and \(r' = h'' r\). Then $u' g + v' s = r'$, $h' u' = \kappa$, $h' v' = \lambda$ and $h' r' = \omega$ for some $h' \in R$.
\end{proposition}

\begin{proof}
Since \(\lambda = h v\), it follows that \(\lclm{x-\alpha_k,v} \mid_r \lambda\), so there exists \(h' \in R\) such that \(\lambda = h' \lclm{x-\alpha_k,v}\). Then \(h v = \lambda = h' h'' v\), hence \(h = h' h''\). Multiplying $u g + v s = r$ by \(h''\) on the left, we get $u' g + v' s = r'$. Moreover, \(\kappa = h u = h' h'' u = h' u'\), \(\lambda = h v = h' h'' v = h' v'\) and \(\omega = h r = h' h'' r = h' r'\). 
\end{proof}

So we need a way to find new right roots of \(\lambda\). 

\begin{proposition}\label{findingnewroot}
Assume \(\lambda = h v\) with \(\deg h \geq 1\). Let \(\{s_1, \dots, s_m\} = \{ i \in \{0, \dots, n-1\} ~|~ (x - \alpha_i) \mid_r v\}\) and \(\{l_1, \dots, l_r\} = \{0, \dots, n-1\} \setminus \{s_1, \dots, s_m\}\). For any $1\leq i\leq r$, let $f_i = \lclm{f_{i-1},x - \alpha_{l_i}}$ with $f_{0} = v$. Then:
\begin{enumerate}
\item There exists \(d \geq 0\) such that $\deg(f_{d-1})=\deg(f_d)$,
\item If \(d_0\) is the minimal index such that $\deg(f_{d_0-1})=\deg(f_{d_0})$, then \(d_0 \in \{k_1, \dots, k_\nu\}\).
\end{enumerate}
\end{proposition}

\begin{proof}
For any $1\leq i\leq r$, let $\lambda_i=\lclm{\lambda_{i-1},x - \alpha_{l_i}}$ with $\lambda_{0} = \lambda$. It is clear that $f_i \mid_r \lambda_i$ for any $1 \leq i \leq r$. Suppose that the sequence $\{\deg(f_i)\}_{0\leq i\leq r}$ is strictly increasing. Hence $\deg(f_{r}) = r + \deg(v) = n - m + \deg(v) > n$ because, by Proposition \ref{criterion}, \(\deg(v) > m\). This is not possible, since $f_{r} \mid_r \lambda_{r} = \lcrm{\{x - \alpha_i ~|~ 0 \leq i \leq n-1\}}$ whose degree is bounded from above by \(n\). So there exists a minimal $d_0 \geq 0$ such that $\deg(f_{d_0-1}) = \deg(f_{d_0})$. Now, $x-\alpha_{i_{d_0}} \mid_r f_{d_0-1} \mid_r \lambda_{d_0-1} = \lclm{\lambda, x-\alpha_{l_1}, \ldots, x-\alpha_{l_{d_0-1}}}$. Since, $l_{d_0} \neq l_1, \ldots, l_{d_0-1}$, $x - \alpha_{l_{d_0}} \mid_r \lambda$. Thus, $d_0 \in \{k_1, \dots, k_\nu\}$.
\end{proof}

Propositions \ref{addingnewroot} and \ref{findingnewroot} provide a way to find the locator if a decoding failure happens. This is presented in Algorithm \ref{decodingfailure}. 

\begin{algorithm}[h]
\caption{Solving decoding failures}\label{decodingfailure}
\begin{algorithmic}[5]
\REQUIRE A skew differential Goppa code \(\mathcal{C}\) of length \(n\), correction capability \(t\), positional points \(\{\alpha_0, \dots, \alpha_{n-1}\} \subseteq L\), \(\eta_0, \dots, \eta_{n-1} \in L^*\), skew differential Goppa invariant polynomial \(g \in R\) with \(\deg(g) = 2t\), and parity check polynomials \(h_0, \dots, h_{n-1} \in R\) of degree \(2t-1\).
\REQUIRE The polynomial \(v_{curr}\) and the set \(pos\) in Algorithm \ref{Sug}
\ENSURE The locator polynomial \(\lambda\).
\WHILE{$\deg(v_{curr}) > |pos|$}\label{decfailure}
	\STATE \(f \gets v_{curr}\), \(e \gets \deg(f)\)
	\STATE \(i \gets \text{one element in } other\), \(other = other \setminus \{i\}\)
	\STATE \(f \gets \lclm{f, x - \alpha_i}\)
	\WHILE{\(\deg(f) > e\)}
		\STATE \(e \gets e + 1\)
		\STATE \(i \gets \text{one element in } other\), \(other = other \setminus \{i\}\)
		\STATE $f\gets \lclm{f, x-\alpha_i}$
	\ENDWHILE
	\STATE \(pos = pos \cup \{i\}\), \(other = \{0, \dots,n-1\} \setminus pos\)
	\STATE \(v \gets v_{curr}\), \(v_{curr} \gets \lclm{v, x-\alpha_i}\), \(h \gets \operatorname{l-quo}(v_{curr},v)\), \(r_{curr} \gets h r_{curr}\)
	\FOR{\(i \in other\)}
		\IF{$\alpha_i$ is a right root of $v_{curr}$}
			\STATE $pos \gets pos \cup \{i\}$, \(other = other \setminus \{i\}\)
		\ENDIF
	\ENDFOR
\ENDWHILE
\RETURN $v_{curr}$
\end{algorithmic}
\end{algorithm}

\section{Parity check matrices and positional points for skew Goppa codes}\label{skewGoppaconstruction}

This section deals with the computation of parity-check matrices and the choice of the positional points for skew Goppa codes. Although most of results are still valid in the skew differential case, the presentation become less technical under the assumption  \(\partial = 0\). On the other hand, this level of generality suffices for our main purpose, namely, the design of a cryptosystem based on skew Goppa codes over a finite field in Section \ref{classicalMcEliece}.

For a given skew differential Goppa code, a parity check matrix can be derived from \eqref{eq:Goppadef}. We make it explicit in the skew Goppa case. So let \(R = L[x;\sigma]\), where \(L\) is a finite extension of a given field \(F\), and \(\sigma\) is a field automorphism of \(L\) of finite order \(\mu\). Let  \(\mathcal{C}\) be a skew Goppa code with Goppa polynomial \(g \in R\), positional points \(\{\alpha_0, \dots, \alpha_{n-1}\}\), \(\eta_0, \dots, \eta_{n-1} \in L^*\) and parity check polynomials \(h_0, \dots, h_{n-1}\).  Let \(\deg(g) = \chi\), \(h_i = \sum_{j=0}^{\chi-1} h_{i,j} x^j\) and 
\[
\widehat{H} = \begin{pmatrix}
\sigma^{-0}(h_{0,0}) \eta_0 & \sigma^{-0}(h_{1,0}) \eta_1 & \cdots & \sigma^{-0}(h_{n-1,0}) \eta_{n-1} \\
\sigma^{-1}(h_{0,1}) \eta_0 & \sigma^{-1}(h_{1,1}) \eta_1 & \cdots & \sigma^{-1}(h_{n-1,1}) \eta_{n-1} \\
\vdots & \vdots & \ddots & \vdots \\
\sigma^{-\chi+1}(h_{0,\chi-1}) \eta_0 & \sigma^{-\chi+1}(h_{1,\chi-1}) \eta_1 & \cdots & \sigma^{-\chi+1}(h_{n-1,\chi-1}) \eta_{n-1} \\
\end{pmatrix}.
\]

\begin{proposition}\label{paritycheckmatrix}
For each \(\gamma \in L\), let \(\tovector(\gamma)\) denote its \(F\)--coordinates, as a column vector, with respect to a fixed \(F\)--basis of \(L\). Let
\[
H = \Big(\begin{matrix} \tovector(\sigma^{-j}(h_{i,j}) \eta_i) \end{matrix}\Big)_{\genfrac{}{}{0pt}{3}{0 \leq j \leq \chi-1}{0 \leq i \leq n-1}} \in \matrixspace{(\chi m) \times n}{F}.
\]
Then \(H\) is a parity check matrix for \(\mathcal{C}\). 
\end{proposition}

\begin{proof}
Observe that
\begin{multline*}
\sum_{i=0}^{n-1} h_i \eta_i c_i = \sum_{i=0}^{n-1} \sum_{j=0}^{\chi-1} h_{i,j} x^j \eta_i c_i = \\
\sum_{i=0}^{n-1} \sum_{j=0}^{\chi-1} x^j \sigma^{-j}(h_{i,j}) \eta_i c_i = \sum_{j=0}^{\chi-1} x^j \sum_{i=0}^{n-1} \sigma^{-j}(h_{i,j}) \eta_i c_i,
\end{multline*}
so \eqref{eq:Goppadef} is also equivalent to 
\[
\sum_{i=0}^{n-1} \sigma^{-j}(h_{i,j}) \eta_i c_i = 0, \ 0 \leq j \leq \chi-1,
\]
i.e.
\[
(c_0, c_1, \dots, c_{n-1}) \transpose{\widehat{H}} = 0.
\]
Since \(\mathcal{C} \subseteq F^{n}\), \((c_0, c_1, \dots, c_{n-1}) \in \mathcal{C}\) if and only if \((c_0, c_1, \dots, c_{n-1}) \transpose{H} = 0\). 
\end{proof}

We gather from \cite{Lam/Leroy:2004, Delenclos/Leroy:2007} the information on P-independent sets needed to describe every possible set of positional points in the skew Goppa case. 

It is well known  that the center of \(R = L[x;\sigma]\) is \(K[x^{\mu}]\), where \(K = L^\sigma\), the invariant subfield of \(L\) under \(\sigma\). So, for every \(a\in L\), the polynomial \(x^\mu - \norm{}{a}\) is central, where 
\[
\norm{}{a} = a\sigma(a) \cdots \sigma^{\mu-1}(a)
\]
is the norm of \(a\). Define, following \cite{Lam/Leroy:1988}, the conjugate of \(a\) under \(c \in L^*\) as \(^{c}a = \sigma(c)ac^{-1} \), and the conjugacy class of \(a \) as 
\[
\Delta(a) = \{ {}^ca : c \in L^*\}. 
\]
By virtue of Hilbert's 90 Theorem (see e.g. \cite[Chapter VI, Theorem 6.1]{Lang:2002}), \(\Delta (a) = \Delta(b)\) if, and only if, \(\norm{}{a} = \norm{}{b}\). Hence,
\begin{equation}\label{Deltanorma}
\Delta(a) = \{b \in L : \norm{}{a} = \norm{}{b}\}.
\end{equation}

Observe that these conjugacy classes form a partition of \(L\).

For each \(f = \sum_j f_j x^j \in R\) and any \(b \in L\), by \cite[Lemma 2.4]{Lam/Leroy:1988}, there exists \(h \in R\) such that 
\begin{equation}\label{leftevaluation}
f = h (x-b) + \sum_{j} f_j \norm{j}{b},
\end{equation}
where \(\norm{j}{b} \in L\) is defined as 
\[
\norm{0}{b} = 1, \text{ and }
\norm{j}{b} = b \sigma(b) \dots \sigma^{j-1}(b) \text{ for }  j \geq 1.
\]
Observe that  \(\norm{}{b} = \norm{\mu}{b}\). We get thus from \eqref{leftevaluation} and \eqref{Deltanorma} that
\begin{equation}\label{Deltadiv}
\Delta(a) = \{ b \in L : (x-b) \mid_r x^\mu - \norm{}{a}\}.
\end{equation}

We are in a position to show how to build P-indepentent sets by using the general theory as stablished in \cite{Lam/Leroy:2004, Delenclos/Leroy:2007}.%

\begin{proposition}\label{boundnumPindep}
Given  \(a \in L^*\), the P-independent subsets of \(\Delta(a)\) are those of the form \(\{{}^{c_1}a, \dots, {}^{c_m}a\}\) where \(m \leq \mu\) and \(\{c_1, \dots, c_m\} \) is a \(K\)--linearly independent subset of \(L\). Moreover, \(m = \mu\) if and only if 
\[
\lclm{x-{}^{c_1}a, \dots, x- {}^{c_m}a} = x^\mu - \norm{}{a}.
\]
\end{proposition}

\begin{proof}
The first statement follows from \cite[Theorem 5.3.iii]{Delenclos/Leroy:2007}, while the second is derived from \eqref{Deltadiv}.
\end{proof}

\begin{proposition}\label{unionPindependent}
A subset \(\Gamma \subseteq L^*\) is P-independent of and only if  \[\Gamma = \Gamma_1 \cup \cdots \cup \Gamma_r,\] where \(\Gamma_i \subseteq \Delta(a_i)\) is P-independent for all \(i=1, \dots, r\), and \(a_1, \dots, a_r \in L\) are nonzero elements of different norm. 
\end{proposition}

\begin{proof}
It follows from \cite[Corollary 4.4]{Lam/Leroy:2004}. 
\end{proof}

As for the selection of the skew Goppa polynomial concerns, we may state:

\begin{proposition}\label{rroots}
Consider \(h \in K[x^\mu] \) without roots in \(K\). Then \(g = x^ah \in L[x;\sigma]\) has no right roots in \(L^*\) for any \(a \geq 0\). 
\end{proposition}
\begin{proof}
If \(\alpha \in L^*\) is a right root of \(g\), then \(\alpha\) is a right root of \(h \in L[x;\sigma]\). Then, by Proposition \ref{boundnumPindep}, \(x^\mu - N(\alpha)\) is a right divisor in \(L[x;\sigma]\) of the central polynomial \(h\). This gives that \(x^\mu - N(\alpha)\) is a divisor of \(h \in K[x^\mu]\), that is, \(N(\alpha) \in K\) is a root of \(h(x^\mu)\). 
\end{proof}

\section{A McEliece cryptosystem based on skew Goppa codes}\label{classicalMcEliece}

To design a skew Goppa code \(\mathcal{C}\), we first choose as alphabet a finite field \(F = \field[q]\) where \(q = p^d\) for a prime \(p\). Next, we set the length \(n\) and the correction capacity \(t < n/2\). In general, this parameter \(t\) will be much smaller than \(n\), as we will see. 

Algorithms \ref{Sug} and \ref{decodingfailure} guarantee that we may set  \begin{equation}\label{tdeg} t = \left\lfloor \frac{\deg g}{2} \right\rfloor, \end{equation}
where \(g\) is the skew Goppa polynomial. So we must built the skew polynomial ring \(R = L[x;\sigma]\), where \(L\) is an extension of \(F\) of degree \(m\), so \(L = \field[q^m] \). We choose \(t, n, m\) such that \(2mt \leq n\) since, from \eqref{eq:Goppadef}, a parity check matrix over \(F\) has size \(2mt \times n\). If \(2mt\) is too close or too far from \(n\), we get codes with very small or very large dimension. For instance, in the Classic McEliece NIST’s Post-Quantum Cryptography Standardization Project proposal, see \cite{ClassicMcEliece:2020}, the proposed code rates, the ratios between dimension and length, are \(\approx 0.75\).

From the relation
\begin{equation}\label{dimrank}
\dim_F \mathcal{C} = n - \operatorname{rank} (H) \geq n - 2mt,
\end{equation}
we obtain that, by choosing 
\[
m \leq \frac{n}{4t},
\]
we get 
\[
\frac{\dim_F \mathcal{C}}{n} \geq 0.5. 
\]

Should the dimension of \(\mathcal{C}\) does not reach the value
\(
 n - 2mt,
\)
then we randomly chose a linear subcode \(\mathcal{C'}\) of \(\mathcal{C}\) with that dimension. If we set 
\[
\frac{n}{10t} \leq m \leq \frac{n}{4t},
\]
then 

\[
0.5 \leq \frac{\dim_F \mathcal{C}'}{n} \leq 0.8. 
\]

The field automorphism \(\sigma\) of \(L\) is given as a power of the Frobenius automorphism \(\tau\), that is \(\tau(a) = a^p\), so we pick \(1 \leq s \leq md\)  and set  \(\delta = \operatorname{gcd}(s,md)\). Define \(\sigma = \tau^s\), which has order
\[
\mu = \frac{dm}{\delta},
\]
and \(K = L^\sigma = \field[p^\delta]\).

The definition of the skew Goppa code \(\mathcal{C}\) requires from the specification of a \(P\)--independent subset of \(L^*\), the positional points, and an invariant polynomial \(g \in R\) having no right root among these points. As for the first task concerns, we will describe all maximal P-independent subsets of \(L^*\). Every other P-independent set is a subset of one of these.
\begin{proposition}\label{normofprimitives}
Let \(\gamma \in L\) a primitive element. Every maximal P-indepedent subset of \(L^*\) if of the form 
\[
\{\sigma(c_{ij})\gamma^ic_{ij}^{-1} : i=0, \dots, p^\delta-{2}, j = 0, \dots, \mu-1\},
\]
where \(\{c_{i0}, \dots, c_{i\mu-1}\}\) is a \(K\)--basis of \(L\) for each \(i=0, \dots, p^\delta-2\). As a consequence, if a  P-independent subset of \(L^*\) has \(n\) elements, then \(n \leq (p^\delta-1)\mu\). 
\end{proposition} 
\begin{proof}
It is well-known that \(N(\gamma)\) is a primitive element of \(K\). Thus, \[\{N(\gamma^i) : i =0, \dots, p^\delta -2 \}\] is a set of representatives of the conjugacy classes of \(L^*\) according to \eqref{Deltanorma}. The proposition holds now from  Propositions \ref{boundnumPindep} and  \ref{unionPindependent}. 
\end{proof}

\begin{example}\label{primitivonormal}
Let \(\{\alpha, \sigma(\alpha), \dots, \sigma^{\mu-1}(\alpha)\}\) be a normal basis of \(L/K\). For \(0 \leq i \leq \mu-1\), set \(\beta_i = \sigma^{i+1}(\alpha)/\sigma^i(\alpha)\). Proposition \ref{normofprimitives} implies that
 \[
\left\{ \gamma^i\beta_j ~|~ 0 \leq i \leq p^\delta-2, 0 \leq j \leq \mu-1 \right\}
\]
is a maximal P-independent set of \(L^*\). 
\end{example}

As for the choice of the skew Goppa polynomial concerns, we may set \(g = x^ah\), for any central non constant polynomial \(h \in K[x^\mu]\) without roots in \(K\) and \(a \geq 0\) (Proposition \ref{rroots}) adjusted to condition \eqref{tdeg}. 

\begin{remark}
If \(g = x^a h\) with \(h\) irreducible we have an isomorphism of rings
\[
\frac{R}{Rg} \cong \frac{R}{Rx^a} \times \frac{R}{Rh}.
\]
The first factor is a non-commutative serial ring of length \(a\), while the second factor is isomorphic to a matrix ring with coefficients in a field extension of \(K\). So, the group of units of \( R/Rg\) is a product of the group of units of a field (if \(a > 0\)) and a general linear group over a field extension of \(K\). 
\end{remark}

By Proposition \ref{normofprimitives} we get the inequality
\begin{equation}\label{nboundedbymdelta}
n \leq \mu(p^\delta - 1) = \frac{dm}{\delta} \left(p^\delta - 1\right). 
\end{equation}
So, given \(n, t, q=p^d\), we want to find \(m, \delta\) such that 
\begin{equation}\label{mdeltarestrictions}
\max \left\{ \frac{n}{10t}, \frac{n \delta}{d(p^\delta - 1)} \right\} \leq m \leq \frac{n}{4t} \text{ and } \delta \mid dm.
\end{equation}

Our proposal of a McEliece cryptosystem follows the dual version of Niederreiter \cite{Niederreiter:1986}, by means of a Key Encapsulations Mechanism like the one proposed in \cite{ClassicMcEliece:2020}. 

\subsection{Key schedule}

The input is \(n \gg t\) and \(F = \field[q]\) with \(q= p^d\).

\subsubsection{Construction of additional parameters}\label{additionalparameters} In order to generate the public and private keys for a McEliece type cryptosystem, the parameters \(m,s\) have to be found. The values of \(m, s\) can be computed randomly via an exhaustive search looking for pairs \(m,\delta\) satisfying \eqref{mdeltarestrictions} and then looking for \(s\) such that \(\delta = \operatorname{gcd}(s,md)\).  For instance, if \(n = 4096, t = 25, q = p^d = 2\), we get the following combinations: 
\[
\begin{array}{|c|cccccccccccc|}
\hline
m & 24 & 26 & 28 & 30 & 32 & 33 & 34 & 36 & 36 & 38 & 39 & 40 \\
\hline
\delta & 12 & 13 & 14 & 15 & 16 & 11 & 17 & 12 & 18 & 19 & 13 & 20 \\
\hline
\end{array}
\]

If \(n = 2560, t = 22, q = p^d = 2^4\), we get \(65\) different combinations, where \(12 \leq m \leq 29\) and \(12 \leq \delta \leq 58\).

We set \(k = n - 2t\left\lfloor \frac{n}{4t} \right\rfloor\), the smallest possible dimension, according to \eqref{mdeltarestrictions}. Next pick randomly \(1 \leq s \leq  dm\), and let \(\delta = \operatorname{gcd}(s,dm)\), \(\mu = \frac{dm}{\delta}\), \(L = \field[q^m]\), \(K = \field[p^\delta]\) and \(\sigma = \tau^s : L \to L\). Fix a basis of \(L\) over \(F\) and denote \(\tovector : L \to F^m\) the map providing the coordinates with respect to this basis. Let also denote \(R = L[x;\sigma]\). 

\subsubsection{Left P-independent set}\label{LeftPindependentset} Our set of positional points may be selected amongst the points in a maximal left P-independent set as computed in Example \ref{primitivonormal}. So we need a normal basis and a primitive element of \(L\).

Let first compute a normal basis of \(L\) over \(K\). First we point out that 
\[
\{\alpha, \sigma(\alpha), \dots, \sigma^{\mu-1}(\alpha)\} = \{\alpha, \tau^{\delta}(\alpha), \dots, \tau^{\delta(\mu-1)}(\alpha)\}
\] 
since both \(\tau^{\delta}\) and \(\sigma\) are generators of the cyclic Galois group of the field extension \(L\) of \(K\). 

For each \(\phi \in K[z]\), let \(\varphi_{p^{\delta}}(\phi)\) be the number of polynomials in the indeterminate \(z\) of degree smaller than \(\deg \phi\) and relatively prime to \(\phi\). It is well known, see \cite[Theorem 3.73]{Lidl/Niederreiter:1997}, that \(\varphi_{p^{\delta}}(z^\mu-1)\) is the number of \(\alpha \in L\) such that \(\{\alpha, \sigma(\alpha), \dots, \sigma^{\mu-1}(\alpha)\}\) is a normal basis. By \cite[Theorem 2]{Frandsen:2000},
\[
\varphi_{p^{\delta}}(z^\mu-1) \geq \frac{p^{\delta\mu}}{e \left\lceil\log_{p^\delta}\mu\right\rceil},
\]
so the probability \(\rho\) of picking and element which generates a normal basis is bounded from below by 
\[
\rho \geq \frac{1}{e \left\lceil\log_{p^\delta}\mu\right\rceil}.
\]

So, a random search must produce soon a normal element. For instance, the probability to randomly choose an element which generates a normal basis in cases \(n = 4096, t = 25, q = p^d = 2\) and \(n = 2560, t = 22, q = p^d = 2^4\) is \(\rho \geq 0.36\).

It remains to provide a fast method to check if an element generates a normal basis. There are quite enough methods to do that in finite fields, see e.g. \cite{vonzurGathen/Giesbrecht:1990,Kaltofen/Shoup:1998},  where randomized algorithms with costs \(\mathcal{O}(\mu^2 + \mu \log p^{\delta})\) and \(\mathcal{O}(\mu^{1.82} \log p^\delta)\) respectively, are provided. In our experiments we have just used the classical Hensel test, see \cite{Hensel:1888} or \cite[Theorem 2.39]{Lidl/Niederreiter:1997}, which says that for a given \(\alpha \in L = \field[p^{dm}]\), \(\{\alpha, \alpha^{p^{\delta}}, \dots, \alpha^{p^{(\mu-1)\delta}}\}\) is a normal basis if and only if
\[
 \operatorname{gcd}\left( z^\mu-1, \alpha z^{\mu-1} + \alpha^{p^{\delta}} z^{\mu-2} + \dots + \alpha^{p^{(\mu-1)\delta}} \right) = 1.
\]

A similar analysis can be done for primitive elements. As mentioned in the introduction of \cite{Shparlinski:2018}, all known algorithms to compute primitive elements work in two steps: compute a reasonable small subset containing a primitive element and test all elements of this subset until a primitive elements is found. Since the number of primitive elements in \(L\) is \(\varphi(|L|-1) = \varphi(p^{dm} - 1)\) and, by \cite[Theorem 328]{Hardy/Wright:1960}, \(\varphi(p^{dm}-1) / (p^{dm}-1)\) is asymptotically bounded from below by a constant multiple of \(\log \log (p^{dm} -1)\), a random search would produce quite fast a primitive element. For instance, in case \(n = 4096, t = 25, q = p^d = 2\) this lower bound is always over \(0.168\), and in case \(n = 2560, t = 22, q = p^d = 2^4\) over \(0.127\).

Testing if a randomly chosen \(\gamma \in L\) is primitive can be done with the classical equivalence
\[
\text{\(\gamma\) is primitive} \iff \gamma^{\frac{p^{dm}-1}{p_i}} \neq 1 \text{ for all prime factor \(p_i\) of \(p^{dm}-1\)}
\]
which requires factoring \(p^{dm}-1\). Since \(p^{dm}-1\) is reasonably small, this can also be done efficiently.   

Once a primitive \(\gamma\) and a normal \(\alpha\) have been randomly computed, a maximal set of left P-independent elements is 
\[
\mathsf{P} = \textstyle\left\{ \gamma^i \frac{\sigma^{j+1}(\alpha)}{\sigma^j(\alpha)} ~|~ 0 \leq i \leq p^\delta-2, 0 \leq j \leq \mu-1 \right\}
\]

\subsubsection{Positional points, skew Goppa polynomial and parity check polynomials}\label{pointsandpolynomials} The list \(\mathsf{E}\) of positional points is obtained by a random selection of \(n\) points in \(\mathsf{P}\). Observe that we have chosen the parameter to have \(n \leq |\mathsf{P}|\). 
\[
\mathsf{E} = \{\alpha_0, \dots, \alpha_{n-1} \} \subseteq \mathsf{P}.
\]

For the skew Goppa polynomial, we randomly choose a monic polynomial \(h(y) \in K[y]\) without roots in \(K\), see Proposition \ref{rroots}, such that \(\deg_y(h) = \left\lfloor 2t / \mu \right\rfloor\) and set \(g = h(x^\mu) x^{2t \bmod \mu}\), which has degree \(2t\). 
 
Finally, the REEA allow to compute \(h_0, \dots, h_{n-1} \in R\) such that, for each \(0 \leq i \leq n-1\), \(\deg(h_i) < 2t\) and 
\[
(x - \alpha_i) h_i - 1 \in Rg.
\]
In fact \(\deg(h_i) = 2t-1\) by a degree argument. 

\subsubsection{Parity check matrix and public key}\label{paritycheckandpublickey} By Proposition \ref{paritycheckmatrix}, a parity check matrix for the skew Goppa code is
\[
H = \Big(\begin{matrix} \tovector(\sigma^{-j}(h_{i,j}) \eta_i) \end{matrix}\Big)_{\genfrac{}{}{0pt}{3}{0 \leq j \leq 2t-1}{0 \leq i \leq n-1}} \in \matrixspace{2tm \times n}{F}
\]
where \(h_i = \sum_{j=0}^{2t-1} h_{i,j} x^j\). 
Once \(H\) is computed, the public key of our cryptosystem can be computed as follows: set \(k = n - 2t\left\lfloor \frac{n}{4t} \right\rfloor\),  \(r_H = \operatorname{rank}(H)\) and \(A \in \matrixspace{(n-k-r_H) \times n}{F}\), a random full rank matrix. The matrix \(H_{\mathrm{pub}}\) is formed by the non zero rows of the reduced row echelon form of the block matrix \(\left(\begin{smallmatrix} H \\ \hline A \end{smallmatrix}\right)\). If \(H_{\mathrm{pub}}\) has less that \(n-k\) rows, pick a new \(A\). This \(H_{\mathrm{pub}}\) defines randomly a linear subcode of \(\mathcal{C}\) of dimension \(k\). 

After this Key Schedule in the Key Encapsulation Mechanism, the different values remain as follows:
\begin{description}
\item[Parameters] \(t \ll n\), \(q = p^d\) and \(k = n - 2t\left\lfloor \frac{n}{4t} \right\rfloor\).
\item[Public Key] \(H_{\mathrm{pub}} \in \matrixspace{(n-k) \times n}{F}\). 
\item[Private Key] \(L\), \(\sigma\), \(\mathsf{E} = \{\alpha_0, \dots, \alpha_{n-1}\}\), \(g\) and \(h_0, \dots, h_{n-1}\). 
\end{description}

\subsection{Encryption procedure: shared key derived by the sender}

We pick a random error vector, i.e. \(e \in F^n\) such that \(\weight(e) = t\), with corresponding error polynomial \(e(x) = \sum_{j=1}^t e_j x^{k_j}\), and \(0 \leq k_1 < k_2 < \cdots < k_t \leq n-1\). The sender can easily derive a shared secret key from \(e\) by means of a fixed and publicly known hash function \(\mathcal{H}\). The cryptogram is
\[
c = e \transpose{H_{\mathrm{pub}}} \in F^{n-k}.
\]

\subsection{Decryption procedure: shared key derived by the receiver}

The receiver can easily compute \(y \in F^n\) such that 
\[
c = y \transpose{H_{\mathrm{pub}}}
\]
since \(H_{\mathrm{pub}}\) is in row reduced echelon form. Let \(y(x) = \sum_{i=0}^{n-1} y_i x^i \in R\). Algorithms \ref{Sug} and \ref{decodingfailure} can be applied to \(y(x)\) in order to compute \(e\). Then the shared secret key can be retrieved by the receiver as \(\mathcal{H}(e)\).

\begin{remark}
Of course the original approach in \cite{McEliece:1978} can also be followed. In this case the public key can be obtained as follows: Compute a full rank generator matrix \(G\), for the left kernel of \(\transpose{H}\), let \(S \in \matrixspace{r \times r}{F}\) a random non singular matrix where \(r = \operatorname{rank}(G)\). Then \(G_{\mathrm{pub}}\) consists in the first \(k\) rows of \(SG\). The encryption procedure starts with a message which is a word \(m \in F^k\). In order to encrypt, we select a random \(e \in F^n\) such that  \(\weight(e) = t\). The cryptogram is
\[
y = m G_{\mathrm{pub}} + e \in F^n.
\]
To decrypt, let \(y(x) = \sum_{i=0}^{n-1} y_i x^i\). The decryption procedure consists in decoding \(y(x)\), which can be done with Algorithms \ref{Sug} and \ref{decodingfailure}, and then the message can be recovered multiplying by a suitable right inverse of \(G_{\mathrm{pub}}\). 
\end{remark}

%\section{Conclusions}
%\label{sec:conclusions}
%
%Some conclusions here.
%

\appendix

\section{A counterexample to Wang's key equation}\label{Wang}

In this appendix, we follow the notation used by Wang in \cite{Wang:2018}. Let \(q = 2\), \(e = 3\), \(m = 9\), \(n = 8\) and \(t = 5\). The fields \(\field[q^e]\) and \(\field[q^m]\) are represented as follows:
\[
\field[2^3] = \field[2][b] / \langle b^3 + b + 1\rangle, \quad \field[2^9] = \field[2][c] / \langle c^9 + c^4 + 1 \rangle
\]
Under these representations, \(b\) and \(c\) are primitive elements. The embedding \(\field[2^3] \subseteq \field[2^9]\) is given by \(b \mapsto c^4 + c^3 + c^2 = c^{292}\). We represent elements in these fields as powers of \(b\) and \(c\) to save space. 

Let 
\begin{multline}\label{eq:gspace}
\mathbf{g} = \left\langle g_1 = c^{479}, g_2 = c^{224}, g_3 = c^{232}, g_4 = c^{158}, \right. \\
\left. g_5 = c^{14}, g_6 = c^{60}, g_7 = c^{121}, g_8 = c^{267} \right\rangle.
\end{multline}
It can be checked that \(\{g_1, \dots, g_8\}\) and \(\{g_1^{-1}, \dots, g_8^{-1}\}\) are linearly independent over \(\field[q] = \field[2]\). 
Let 
\[
L = x^{2^5} + c^{15} x^{2^4} + c^{414} x^{2^3} + c^{413} x^{2^2} + c^{53} x^{2^1} + c^{377} x^{2^0},
\]
a linearized polynomial. Then \(L = r_i x\ {}_L\mathrm{mod}\ x^2 - g_i^{2-1} x\) where \(r_1 = c^{454}, r_2 = c^{194}, r_3 = c^{402}, r_4 = c^{117}, r_5 = c^{297}, r_6 = c^{353}, r_7 = c^{90}, r_8 = c^{178}.\)

In order to define the code \(\mathcal{LG}(\mathbf{g},L(x))\), inverses \((x^q - g_i^{q-1}x)^{-1}\) have to be computed. However there are two possible meanings for this inverse, it is the unique linearized polyomial such that 
\[
(x^q - g_i^{q-1}x)^{-1} \circ (x^q - g_i^{q-1}x) = 1\ \mathrm{mod}_R\ L(X)
\]
or such that
\begin{equation}\label{eq:inverse}
(x^q - g_i^{q-1}x) \circ (x^q - g_i^{q-1}x)^{-1}{} = 1\ {}_L\mathrm{mod}\ L(X),
\end{equation}
which are different unless \(L(x)\) is invariant. 
In \cite{Wang:2018} it is not explained which one is chosen. However, it seems that \eqref{eq:inverse} is the one used later on. So we guess this is the right assumption. Thus
\begin{align*}
(x^2 - g_1^{2-1} x)^{-1} &= c^{59} x^{2^4} + c^{164} x^{2^3} + c^{334} x^{2^2} + c^{280} x^{2^1} + x^{384} x^{2^0} \\
(x^2 - g_2^{2-1} x)^{-1} &= c^{296} x^{2^4} + c^{336} x^{2^3} + c^{410} x^{2^2} + c^{505} x^{2^1} + c^{189} x^{2^0} \\
&\vdots
\end{align*}
We are not going to use the remaining inverses in this counterexample, but they can also be computed in a similar way. 

Assume the following error has been introduced during a transmission,
\[
\mathbf{e} = \left( b, b^6, 0, 0, 0, 0, 0, 0 \right) \in {\field[q^e]}^8 = {\field[2^3]}^8.
\]
Then \(w_h(\mathbf{e}) = 2 < \frac{t}{2} = \frac{5}{2}\) and \(M = \{1,2\}\). We can see \(\mathbf{e}\) as a vector over \(\field[q^m]\) via the embedding \(b \mapsto c^{292}\), so 
\[
\mathbf{e} = \left( c^{292}, c^{219}, 0, 0, 0, 0, 0, 0 \right) \in {\field[q^m]}^8 = {\field[2^9]}^8.
\]
Therefore
\begin{align*}
S(x) &= \sum_{i=1}^n (x^q - g_i^{q-1}x)^{-1} \circ e_i x \\
&= \left(c^{59} x^{2^4} + c^{164} x^{2^3} + c^{334} x^{2^2} + c^{280} x^{2^1} + x^{384} x^{2^0}\right) c^{292} x \\
&\quad + \left( c^{296} x^{2^4} + c^{336} x^{2^3} + c^{410} x^{2^2} + c^{505} x^{2^1} + c^{189} x^{2^0} \right) \circ c^{219} x \\ 
&= c^{226} x^{2^4} + c^{165} x^{2^3} + c^{378} x^{2^2} + c^{149} x^{2^1} + c^{436} x^{2^0}, \\
\sigma(x) &= \prod_{g \in \langle g_i: i\in M\rangle} (x - g) = \prod_{g \in \langle g_1, g_2 \rangle} (x - g)\\
&= x^{2^2} + c^{81} x^{2^1} + c^{225} x^{2^0}, \\
\omega(x) &= \sum_{i \in M} e_i \prod_{g \in \langle g_j: j \neq i, j\in M\rangle} (x - g) \\
&= c^{438} x^{2^1} + c^{193} x^{2^0}.
\end{align*}
Hence
\[
\sigma(x) \circ S(x) \ \mathrm{mod}_R\ L(x) = c^{507} x^{2^4} + c^{76} x^{2^3} + c^{32} x^{2^2} + c^{301} x^{2^1} + c^{450} x^{2^0},
\]
which is not equal to \(\omega(x)\). This example shows that the key equation in \cite{Wang:2018} is not always satisfied. The computations were mad with the help of \cite{sage}, using the isomorphism
\begin{equation*}
\begin{split}
\mathcal{L}_q(x,\field[q^m]) &\to \field[q^m][y,\tau_q] \\
\sum_i c_i x^{q^i} &\mapsto \sum_i c_i y^i, 
\end{split}
\end{equation*}
where \(\tau_q\) is the \(q\)-Frobenius automorphism, i.e. \(\tau_q(a) = a^q\).

Beyond the remark we did before concerning the meaning of \((x^q - g_i^{q-1}x)^{-1}\), there are several mistakes in \cite{Wang:2018} which, in our opinion, are the cause of the inaccuracy of his key equation. The first one is \cite[Proposition 1]{Wang:2018}. It is true that \(\sigma_{\mathbf{g}}(x)\) is a left multiple of \(x^q - g_i^{q-1}x\) for all \(1 \leq i \leq n\), in fact it is the least common left multiple of them. However it is not true that \(\sigma_{\mathbf{g}}(x) = \sigma_{\mathbf{g}_i}(x) \circ (x^q - g_i^{q-1}x)\). In fact, for linearly independent \(g_1, g_2\) previous equation becomes 
\[
\sigma_{\left\langle g_1, g_2 \right\rangle}(x) = (x^q - g_2^{q-1} x) \circ (x^q - g_1^{q-1} x).
\]
Taking \(g_1, g_2\) from \eqref{eq:gspace}, we get
\[
\sigma_{\left\langle g_1, g_2 \right\rangle}(x) = x^{2^2} + c^{81} x^{2^1} + c^{225} x^{2^0}, 
\] 
but 
\begin{multline*}
(x^q - g_2^{q-1} x) \circ (x^q - g_1^{q-1} x) = \\
(x^2 - (c^{224})^{2-1} x) \circ (x^2 - (c^{479})^{2-1} x) = x^{2^2} + c^{346} x^{2^1} + c^{456} x^{2^0} \neq \sigma(x)
\end{multline*}

For each \(i \in M\),
\(
\prod_{g \in \langle g_j: j \neq i, j\in M\rangle} (x - g)
\)
should be replaced by the linearized polynomial \(\rho_i(x)\) such that 
\[
\sigma(x) = \prod_{g \in \langle g_i: i \in M\rangle} (x - g) = \rho_i(x) \circ (x^q - g_i^{q-1}x)
\]
in the definition of the evaluator polynomial \(\omega(x)\). But even in this case
\begin{equation*}
\begin{split}
\sigma(x) \circ S(x) &= \sum_{i=1}^n \sigma(x) \circ (x^q - g_i^{q-1}x)^{-1} \circ e_ix \ \mathrm{mod}_R L(x) \\
&= \sum_{i=1}^n \rho_i(x) \circ (x^q - g_i^{q-1}x) \circ (x^q - g_i^{q-1}x)^{-1} \circ e_ix \ \mathrm{mod}_R L(x), 
\end{split}
\end{equation*}
thus \(L(x)\) should be invariant and the evaluator polynomial should be defined as 
\[
\omega(x) = \sum_{i=1}^n \rho_i(x) \circ e_ix
\]
in order to get the key equation. Under these assumptions, the key equation becomes an instance of \eqref{keyeqmod}.

\section{How unlikely are decoding failures?}

First, observe that, by Theorem \ref{thkeyeq} and Proposition \ref{criterion}, decoding failure cannot happen if \(\gcld{\lambda, \omega} = 1\). Next, we analyze this condition.

\begin{proposition}\label{cyclicgenerator}
Under the notation of Section \ref{skewpols}, the following statements are equivalent:
\begin{enumerate}
\item \(\gcld{\omega,\lambda} = 1\).
\item \(\omega + \lambda R\) generates \(R/\lambda R\) as a right \(R\)--module.
\item The set  $\{ (\omega + \lambda R)x^i ~|~ 0 \leq i \leq \nu-1 \}$ is right linearly independent over $L$. 
\end{enumerate}
\end{proposition}

\begin{proof}
The equivalence between (1) and (2) is a direct consequence of Bezout's Theorem. It is clear that $\omega + \lambda R$ generates the right $R$--module $R/\lambda R$  if and only if  $\{ (\omega + \lambda R)x^i ~|~ 0 \leq i \leq \nu-1 \}$ spans $R/\lambda R$ as a right $L$--vector space. Since the dimension over $L$ of $R/\lambda R$ is $\nu$, the equivalence between (2) and (3) becomes clear. 
\end{proof}

In the skew case, i.e. \(\partial = 0\), a more precise analysis can be done. Besides  the partial norms \(N_i(a)\), for any \(a \in L\) and \(i \in \Nset\), the \(-i\)th norm of \(a\) is defined as \(\norm{-i}{a} = \norm[\sigma]{-i}{a} = \norm[{\sigma^{-1}}]{i}{a}\), i.e.
\[
\norm{-i}{a} = a \sigma^{-1}(a) \dots \sigma^{-i+1}(a).
\]
Since \(\sigma\) has order \(\mu\), it follows that \(\norm{\mu}{\gamma} = \norm{-\mu}{\gamma}\). Moreover, for each \(f = \sum_j f_j x^j \in R\) and all \(\gamma \in L\), there exists \(h \in R\) such that 
\begin{equation}\label{rightevaluation}
f = (x-\gamma) h + \sum_{j} \sigma^{-j}(f_j) \norm{-j}{\gamma}.
\end{equation}

\begin{lemma}\label{coordinates}
The $j$-coordinate of \(\omega x^i + \lambda R\) with respect to the basis $\{\rho_{k_1}, \dots, \rho_{k_\nu}\}$ is $\norm{-i}{\alpha_{k_j}} \sigma^{-i}(\eta_{k_j}) \sigma^{-i}(e_j)$, for any $1\leq j \leq \nu$.
\end{lemma}

\begin{proof}
By \eqref{rightevaluation}, $\alpha_{k_j}$ is a left root of $x^i-\norm{-i}{\alpha_{k_j}}$. Then $x^i-\norm{-i}{\alpha_{k_j}} \in (x - \alpha_{k_j}) R$. Multiplying on the left by $\rho_{k_j}$, $\rho_{k_j} x^i - \rho_{k_j} \norm{-i}{\alpha_{k_j}} \in \lambda R$. Thus, in $R/\lambda R$,
\begin{equation*}
\begin{split}
\omega x^i &= \sum_{j=1}^\nu \rho_{k_j} \eta_{k_j} e_j x^i \\ 
&= \sum_{j=1}^\nu \rho_{k_j} x^i \sigma^{-i}(\eta_{k_j}) \sigma^{-i}(e_j) \\
&= \sum_{j=1}^\nu \rho_{k_j} \norm{-i}{\alpha_{k_j}} \sigma^{-i}(\eta_{k_j}) \sigma^{-i}(e_j)
\end{split}
\end{equation*}
and the result follows.
\end{proof}

\begin{proposition}
\(\gcld{\omega,\lambda} = 1\) if and only if
\[
\det \Big( \norm{-i}{\alpha_{k_j}} \sigma^{-i}(\eta_{k_j}) \sigma^{-i}(e_j) \Big)_{0 \leq i \leq \nu-1 \atop 1 \leq j \leq \nu} \neq 0.
\]
\end{proposition}

\begin{proof}
It follows from Lemma \ref{coordinates} and Proposition \ref{cyclicgenerator}.
\end{proof}

\begin{example}
This example shows that decoding failures, although quite unusual, can happen. The computations were done with the help of the SAGEMATH system \cite{sage}.

Let \(L = F = \field[2^8] = \field[2][z]/\left\langle z^8 + z^4 + z^3 + z^2 + 1 \right\rangle\), \(n = 16\), \(t = 2\). Choose the automorphism \(\sigma : L \to L\) defined by \(\sigma(a) = a^{2^4}\). So, \(m = 1\), \(\delta = 4\) and \(\mu = 2\). We have \(K = \field[2^4]\), which may be presented as \(K = \field[2][w] / \left\langle w^4 + w + 1 \right\rangle\), with embedding \(w \mapsto z^{34}\) into \(F\). The chosen normal and primitive elements are \(\alpha = z^{37}\) and \(\gamma = z^{41}\). 

The list \(\mathsf{E} = \{\alpha_0, \dots, \alpha_{15}\}\) of evaluation points contains the elements
\begin{align*}
\alpha_0 &= \gamma^0 \textstyle\frac{\sigma(\alpha)}{\alpha} = z^{45}, &
\alpha_1 &= \gamma^9 \textstyle\frac{\sigma(\alpha)}{\alpha} = z^{159}, \\
\alpha_2 &= \gamma^{13} \textstyle\frac{\sigma(\alpha)}{\alpha} = z^{68}, & 
\alpha_3 &= \gamma^{13} \textstyle\frac{\sigma^2(\alpha)}{\sigma(\alpha)} = z^{233}, \\ 
\alpha_4 &= \gamma^{10} \textstyle\frac{\sigma^2(\alpha)}{\sigma(\alpha)} = z^{110}, & 
\alpha_5 &= \gamma^7 \textstyle\frac{\sigma(\alpha)}{\alpha} = z^{77}, \\ 
\alpha_6 &= \gamma^{12} \textstyle\frac{\sigma(\alpha)}{\alpha} = z^{27}, & 
\alpha_7 &= \gamma^{10} \textstyle\frac{\sigma(\alpha)}{\alpha} = z^{200}, \\ 
\alpha_8 &= \gamma^2 \textstyle\frac{\sigma^2(\alpha)}{\sigma(\alpha)} = z^{37}, & 
\alpha_9 &= \gamma^0 \textstyle\frac{\sigma^2(\alpha)}{\sigma(\alpha)} = z^{210}, \\ 
\alpha_{10} &= \gamma^6 \textstyle\frac{\sigma^2(\alpha)}{\sigma(\alpha)} = z^{201}, & 
\alpha_{11} &= \gamma^3 \textstyle\frac{\sigma(\alpha)}{\alpha} = z^{168}, \\ 
\alpha_{12} &= \gamma^{11} \textstyle\frac{\sigma^2(\alpha)}{\sigma(\alpha)} = z^{151}, & 
\alpha_{13} &= \gamma^2 \textstyle\frac{\sigma(\alpha)}{\alpha} = z^{127}, \\ 
\alpha_{14} &= \gamma^1 \textstyle\frac{\sigma^2(\alpha)}{\sigma(\alpha)} = z^{251}, & 
\alpha_{15} &= \gamma^{12} \textstyle\frac{\sigma^2(\alpha)}{\sigma(\alpha)} = z^{192}.
\end{align*}

Let \(g = x^4 + z^{238}x^2 + z^{68}\) be the skew Goppa polynomial. The corresponding parity check polynomials are
\begin{align*}
h_0 &= z^{136}x^3 + z^{91}x^2 + z^{187}x + z^{142}, &
h_1 &= z^{68}x^3 + z^{62}x^2 + z^{136}x + z^{130}, \\
h_2 &= z^{102}x^3 + z^{170}x^2 + z^{204}x + z^{17}, &
h_3 &= z^{102}x^3 + z^{5}x^2 + z^{204}x + z^{107}, \\
h_4 &= z^{85}x^3 + z^{60}x^2 + z^{34}x + z^{9}, &
h_5 &= z^{238}x^3 + z^{195}x^2 + z^{204}x + z^{161}, \\
h_6 &= z^{85}x^3 + z^{7}x^2 + z^{170}x + z^{92}, &
h_7 &= z^{85}x^3 + z^{225}x^2 + z^{34}x + z^{174}, \\
h_8 &= z^{170}x^3 + z^{252}x^2 + z^{187}x + z^{14}, &
h_9 &= z^{136}x^3 + z^{181}x^2 + z^{187}x + z^{232}, \\
h_{10} &= z^{102}x^3 + z^{3}x^2 + z^{238}x + z^{139}, &
h_{11} &= z^{136}x^3 + z^{19}x^2 + z^{136}x + z^{19}, \\
h_{12} &= z^{170}x^3 + z^{36}x^2 + z^{34}x + z^{155}, &
h_{13} &= z^{170}x^3 + z^{162}x^2 + z^{187}x + z^{179}, \\
h_{14} &= z^{51}x^3 + z^{242}x^2 + z^{221}x + z^{157}, &
h_{15} &= z^{85}x^3 + z^{97}x^2 + z^{170}x + z^{182}.
\end{align*}
From these parity check polynomials the public key can be computed, we obtain
\[
H_{\mathrm{pub}} = 
\begin{scriptsize}
\left(
\begin{array}{cccccccccccccccc}
1 & 0 & 0 & 0 & 0 & 0 & 0 & 0 & z^{142} & z^{92} & z^{126} & z^{156} & z^{187} & z^{178} & z^{234} & z^{88} \\
0 & 1 & 0 & 0 & 0 & 0 & 0 & 0 & z^{73} & z^{103} & z^{157} & z^{113} & z^{188} & z^{253} & z^{222} & z^{152} \\
0 & 0 & 1 & 0 & 0 & 0 & 0 & 0 & z^{109} & z^{109} & z^{64} & z^{165} & z^{131} & z^{204} & z^{138} & z^{145} \\
0 & 0 & 0 & 1 & 0 & 0 & 0 & 0 & z^{180} & z^{78} & z^{202} & z^{230} & z^{82} & z^{81} & z^{185} & z^{224} \\
0 & 0 & 0 & 0 & 1 & 0 & 0 & 0 & z^{70} & z^{247} & z^{51} & z^{65} & z^{49} & z^{162} & z^{111} & z^{36} \\
0 & 0 & 0 & 0 & 0 & 1 & 0 & 0 & z^{119} & z^{236} & z^{50} & z^{243} & z^{136} & z^{56} & z^{133} & z^{225} \\
0 & 0 & 0 & 0 & 0 & 0 & 1 & 0 & z^{89} & z^{172} & z^{152} & z^{209} & z^{234} & z^{22} & z^{231} & z^{96} \\
0 & 0 & 0 & 0 & 0 & 0 & 0 & 1 & z^{70} & z^{152} & z^{157} & z^{32} & z^{247} & z^{180} & z^{172} & z^{106}\end{array}
\right)
\end{scriptsize}
\]
Let the error vector be 
\[
e = \left(z^{249}, 0, 0, 0, 0, 0, 0, 0, 0, 1, 0, 0, 0, 0, 0, 0\right).
\]
Hence the cryptogram is
\[
c = e \transpose{H_{\mathrm{pub}}} = \left(z^{133}, z^{103}, z^{109}, z^{78}, z^{247}, z^{236}, z^{172}, z^{152})\right),
\]
which is transmitted to the receiver. A solution of 
\[
c = y \transpose{H_{\mathrm{pub}}} 
\]
is 
\[
y = \left(z^{133}, z^{103}, z^{109}, z^{78}, z^{247}, z^{236}, z^{172}, z^{152}, 0, 0, 0, 0, 0, 0, 0, 0\right),
\]
which allows to compute the syndrome polynomial 
\[
s = z^{36}x^3 + z^{81}x^2 + z^{87}x + z^{132}.
\]
The LEEA applied to \(g,s\) until we find the first remainder with degree below \(2\), and we get
\[
v_I = z^{189}x + z^{174}, \ r_I = z^{119}.
\] 
The only right  root of \(v_I\) is \(z^{240}\), which is not in \(\mathsf{E}\). Therefore, the cardinal of the evaluation points which are roots of \(v_I\) is \(0 < 1 = \deg v_I\). There is a decoding failure which we can solve with Algorithm \ref{decodingfailure}. The tenth evaluation point, \(\alpha_9\), does not increment the degree, so it is a root of the locator polynomial \(\lambda\). Since \(\lclm{v_I, x - \alpha_9} = x^2 + 1\) which have \(\alpha_0\) and \(\alpha_9\) as roots, we get that \(\lambda = x^2 + 1\). The corresponding evaluator  polynomial is \(\omega = z^{155} x + z^{200}\). The error positions are \(0\) and \(9\), and the evaluator polynomial allows to compute the error values, \(z^{249}\) and \(1\) respectively, as expected. 
\end{example}

%\section*{Acknowledgments}

%\bibliographystyle{siamplain}
%\bibliography{references}

\end{document}